\documentclass[11pt,twoside]{article}
\usepackage{asp2010}

\resetcounters

\begin{document}

\title{Broad Absorption Line Variability in Radio-Loud Quasars}

\author{B.~P.~Miller,$^{1}$ C.~A.~Welling,$^{2}$ W.~N.~Brandt,$^{3,4}$ 
  and R.~R.~Gibson$^{5}$
\affil{$^1$Dept.~of Astronomy, Univ.~of Michigan, 500
  Church Street, Ann Arbor, MI}
\affil{$^2$Dept.~of Physics and Astronomy, Dickinson
  College, Carlisle, PA}
\affil{$^3$Dept.~of Astronomy and Astrophysics, The
  Pennsylvania State University, 525 Davey Laboratory, University
  Park, PA}
\affil{$^4$Institute for Gravitation and the Cosmos, Penn State,
  University Park, PA}
\affil{$^5$Dept.~of Astronomy, Univ.~of Washington,
  Box 351580, Seattle, WA}}

\begin{abstract}
We present preliminary results from an investigation into broad
absorption line (BAL) variability within a sample of 41 radio-loud
quasars (RLQs). Using 28 new Hobby-Eberly Telescope (HET) spectra
along with earlier Sloan Digital Sky Survey (SDSS) or other archival
data, we generate a total set of 50 pairs of BAL equivalent width
measurements. Absorption variability in BAL RLQs typically consists of
modest changes in the depth of trough segments, and variability is
more common on longer rest-frame timescales; these tendencies are
similar to previous findings for BAL radio-quiet quasars (RQQs). BAL
variability in RLQs does not show any obvious dependence upon radio
luminosity or loudness, but there is suggestive support for greater
fractional variability within lobe-dominated RLQs.
\end{abstract}

We have conducted the first systematic survey of BAL variability
within RLQs (Welling et al., in prep). BAL RLQs were identified
through cross-matching SDSS BAL quasar catalogs (Trump et al.~2006;
Gibson et al.~2009) to the FIRST radio survey. New HET targets were
chosen to cover a wide range in radio and \ion{C}{iv} BAL properties
(note that the sample is predominantly composed of high-ionization BAL
quasars). HET observations were carried out in 2007--2008 and 2011 in
queue-scheduled mode. The Low-Resolution Spectrograph was used with
the g2 grating and a 1.5$''$ slit, providing a spectral resolution of
$R\simeq$870 (sufficient for comparison to SDSS spectra). Data were
reduced within {\it IRAF\/} following standard methods.

We find that \ion{C}{iv} BAL variability in RLQs, where present,
generally involves modest changes in the depth of trough segments. BAL
variability was quantified using the absolute change and fractional
change in equivalent width ($|{\Delta}EW|$ and
$|{\Delta}EW/{\langle}EW{\rangle}|$). BAL RLQs tend to vary more on
longer rest-frame timescales: there is a significant correlation
between $|{\Delta}EW/{\langle}EW{\rangle}|$ and ${\Delta}\tau$, and
the mean $|{\Delta}EW/{\langle}EW{\rangle}|$ is $0.17\pm0.04$
($0.07\pm0.01$) for ${\Delta}\tau>500$~d ($<500$~d). Similar
tendencies have been established for BAL RQQs (Gibson et al.~2010;
Capellupo et al.~2011). We construct a comparison sample of BAL RQQs
from Barlow (1993), Lundgren et al.~(2007), and Gibson et
al.~(2010). KS tests indicate that the distribution of BAL variability
indicators (Figure~1, left) is not inconsistent ($p>0.1$) between RLQs
and a ${\Delta}\tau$-matched subset of RQQs (after filtering out
objects with ${\langle}EW{\rangle}<3.5$~\AA). BAL variability in RLQs
appears similar to that in RQQs, supportive of a common physical
mechanism of BAL production.

Radio luminosity (${\ell}_{\rm r}$) and radio loudness ($R^{*}$) do
not appear to influence BAL variability strongly; neither
$|{\Delta}EW|$ nor $|{\Delta}EW/{\langle}EW{\rangle}|$ are
significantly correlated with either ${\ell}_{\rm r}$ or
$R^{*}$. Lobe-dominated RLQs may tend toward greater fractional
variability (Figure~1, right); even omitting three long-timescale
measurements for the lobe-dominated BAL RLQ PG~1004+130, the mean
$|{\Delta}EW/{\langle}EW{\rangle}|$ is $0.23\pm0.07$ ($0.07\pm0.01$)
for lobe-dominated (core-dominated) RLQs. Due to the small number of
lobe-dominated BAL RLQs in our sample and their generally greater
${\Delta}\tau$ and smaller ${\langle}EW{\rangle}$ values, further
study is warranted. Nonetheless, these results may support some
geometrical dependence to BAL structure in RLQs, as is also indicated
by the generally steeper radio spectral indices of BAL RLQs recently
found by DiPompeo et al.~(2011).

\begin{figure}
\epsscale{1.0}
\plottwo{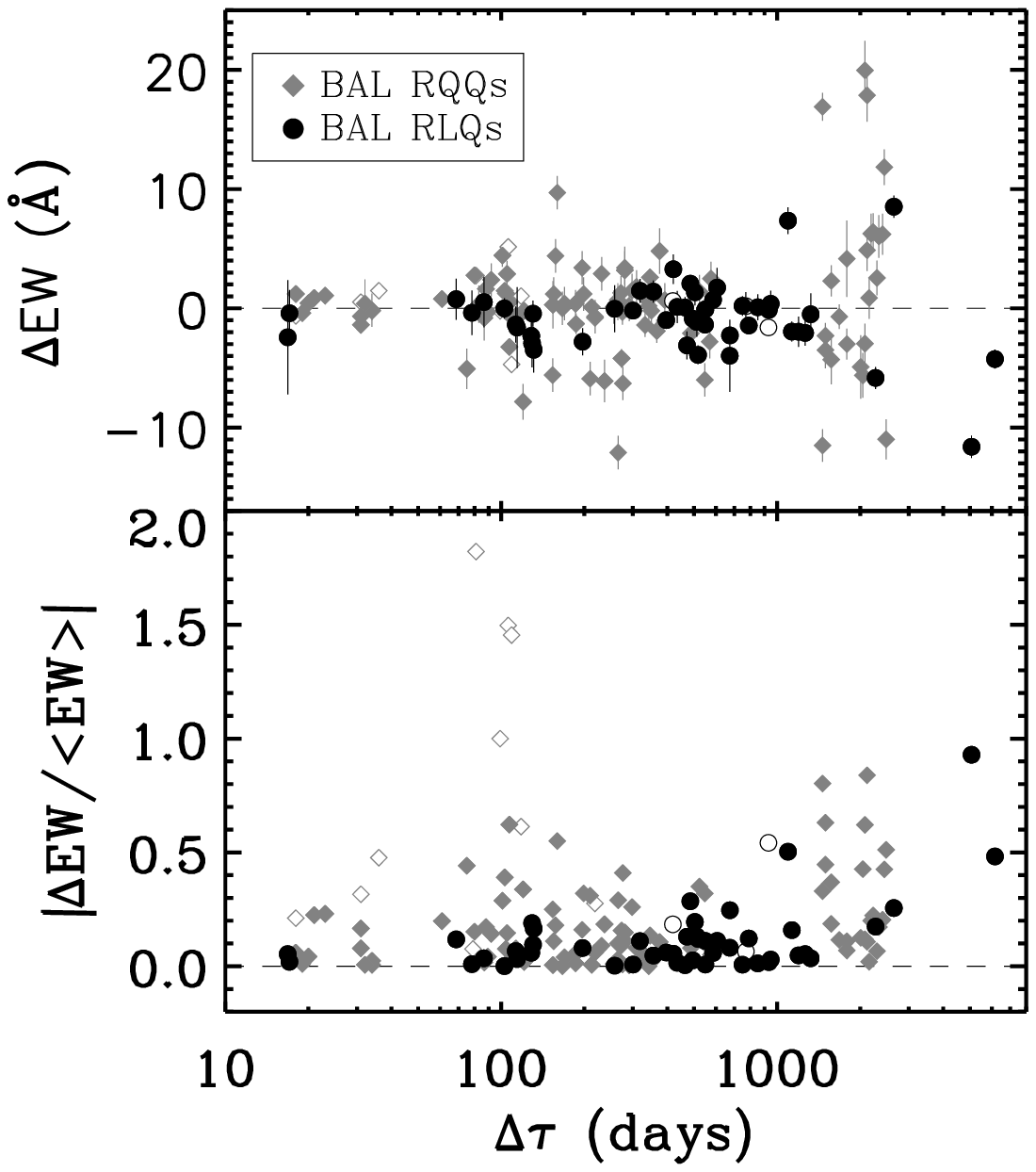}{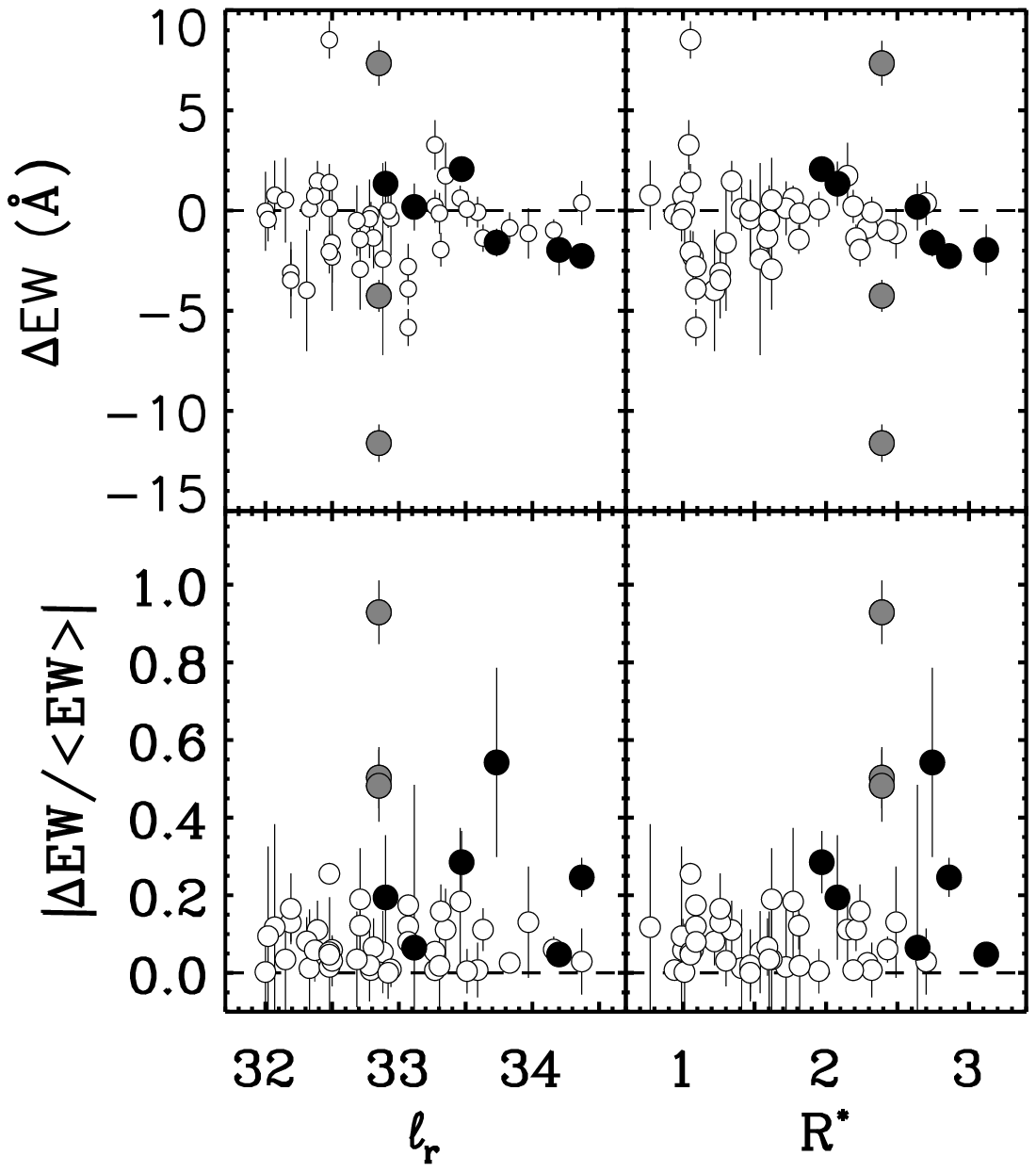}
\caption{{\it Left\/}: BAL variability versus rest-frame timescale
between observations for RQQs (gray diamonds) and RLQs (black
circles). Open symbols have \hbox{$EW<3.5$\AA}. {\it Right\/}: BAL
variability in RLQs versus radio luminosity and radio loudness for
core-dominated (open) and lobe-dominated (filled) objects. The three
long-timescale values for the lobe-dominated PG~1004+130 are plotted
in gray.}
\end{figure}

\acknowledgements We thank Mike Eracleous and Karen Lewis for discussions. 


\end{document}